\documentclass[aps,prb,floatfix,twocolumn,showpacs]{revtex4}
\usepackage{epsfig,amsmath,amssymb,rotating,multirow}

\newcommand{\ba}{\begin{eqnarray}}
\newcommand{\ea}{\end{eqnarray}}

\def\ben{\begin{equation}}
\def\een{\end{equation}}
\def\bea{\begin{eqnarray}}
\def\eea{\end{eqnarray}}
\def\be{\begin{equation}}
\def\ee{\end{equation}}
\def\nn{\nonumber}

\begin{document}

\hsize\textwidth\columnwidth\hsize\csname@twocolumnfalse\endcsname

\title{Effects of Interface Steps on the Valley Orbit coupling in a Si/SiGe quantum dot }

\author{Bilal Tariq and Xuedong Hu}
\affiliation{ Department of Physics, University at Buffalo, SUNY, Buffalo, New York 14260, USA}

\vskip3.5truecm
\begin{abstract}
Valley-orbit coupling is a key parameter for a silicon quantum dot in determining its suitability for applications in quantum information processing.  In this paper we study the effect of interface steps on the magnitude and phase of valley-orbit coupling for an electron in a silicon quantum dot.  Within the effective mass approximation, we find that the location of a step on the interface is important in determining both the magnitude and the phase of the valley-orbit coupling in a Si/SiGe quantum dot.  Specifically, our numerical results show that the magnitude of valley orbit coupling can be suppressed up to 75\% by a step of one atomic monolayer, and its phase can change by almost $\pi$.  When two steps are present, the minimum value of the valley-orbit coupling can even approach zero.  We also clarify the effects of an applied external magnetic field and the higher orbital states on the valley-orbit coupling.  Overall, our results illustrate how interface roughness affect the valley-orbit coupling in silicon, and how spin qubits in silicon may be affected.
\end{abstract}

\maketitle

\section{Introduction }
Silicon is an ideal host for electron or nuclear spin qubits because of its low abundance of spinful isotopes (which can be further reduced through isotopic enrichment) \cite{Lyon_NatureReview, Zanenberg_RMPReview}.  Extremely long coherence times for both electron and nuclear spins have been reported in bulk samples back in the 1950s \cite{Feher_PR1955}, and more recently observed in Si nanostructures for single spins \cite{Morello_Nature2010, Pla_Nature2012, Muhonen_NatNano2014, Veldhorst_NatNano2014}.  There have been multiple experimental demonstration of high-fidelity quantum gates for single-spin qubits \cite{Takeda_SciAdv, Yoneda, Petta_2019,Petta_2018, Petta1, Petta2, Petta3, PettaAPL} and encoded multi-spin systems \cite{HRL,HRL_2, Friesen_2007,Friesen_2006,Friesen_2010,Friesen_2013,Friesen_2015,Friesen_2015_HighFiedelity,Friesen_2018}. The impressive quantum coherence, together with the vast semiconductor industry and its ever improving technological base, makes spin qubits in Si an attractive and promising building block for a future scalable quantum computer.

Significant technological challenges remain toward a fault-tolerant spin qubit in Si, from controlling the effects of charge noise on two-qubit gates to fast and high-fidelity spin measurement.  A particular issue of current interest is the consequences of the multi-valley structure of the Si conduction band \cite{Ando_RMP, Lyon_NatureReview, Zanenberg_RMPReview}.  Conduction band in bulk Si has six degenerate minima.  For electrons confined at an interface four of the valleys have elevated energy so that only two are left in the low-energy sector of the Hilbert space.  Scattering at the interface further couples the last two valleys (valley-orbit coupling) and lifts the remaining degeneracy.  However this valley splitting tends to be relatively small, about 0.2 to 0.3 meV at a Si/SiO$_2$ interface \cite{Dzurak_a,Dzurak_b} and up to 0.1 meV at a Si/SiGe interface \cite{Friesen_2007, Friesen_2006, Friesen_2010, Friesen_2013, Friesen_2015, Friesen_2015_HighFiedelity,Friesen_2018, Petta_2018, Petta1, Petta2, Petta3}, which is usually further reduced by any interface roughness \cite{ PettaAPL, Culcer_2010a, Culcer_2010b}.

The presence of low-energy orbital excited states in a quantum dot could potentially weaken the foundation for a spin qubit and render it susceptible to leakage and decoherence.  Therefore existing experimental and theoretical studies of valley effects have mostly focused on the valley splitting \cite{Saraiva_2009, Saraiva_2011, Friesen_2007, Friesen_2006, Friesen_2010, Friesen_2013, Friesen_2015, Friesen_2015_HighFiedelity, Friesen_2018, Dzurak_a,Dzurak_b,  Petta_2018, Huang, Huang_b, Culcer_2010a, Culcer_2010b, TB_calculations_a, TB_calculations_a_1, TB_calculations_b, TB_2006, TB_2008,TB_calculations_2018, TB_calculations_2018France}, with particular interest in how atomic scale features at the interface affect the valley splitting.  However, it is important to note that valley-orbit (VO) coupling is in general complex, with its magnitude giving the valley splitting while its phase determining how the valleys are mixed at the interface. In a multi-dot system both are important to the electron spectra, with the latter a determining factor for all the tunneling matrix elements \cite{Friesen_2013, Huang2017, Zhao_Sipreprint}.  As we explore multi-dot systems, full knowledge of the valley-orbit coupling would be crucial for a complete understanding of the spin and orbital dynamics.

In this paper we study how one or two well-defined interface steps affect both the magnitude and phase of the valley-orbit coupling within the effective mass approximation.  With a single interface step, we find that magnitude of the VO coupling can be reduced by up to 75\%, while its phase can vary up to $\sim \pi$.  When two interface steps are present and are strategically located, the magnitude of the VO coupling can even vanish completely, while the phase of the VO coupling can vary dramatically depending on where the steps are. In short, our results show that both the magnitude and the phase of the VO coupling are sensitive to the location of the interface step(s).  It is thus inevitable that different quantum dots on the same interface would have different VO couplings if interface steps are present.  Such variability in the VO coupling could significantly impact two-dot properties that are important to electron-based quantum information processing, such as tunnel coupling and exchange coupling.

The rest of the paper is organized as follows. In Section II we present the theoretical framework of our system and defines the Hamiltonians and wavefunctions of our models. In Sec. \ref{VO_section}, we calculate the VO coupling for the ground valley states and analyze our results for different geometries described in the previous section. Section \ref{B_section} covers the effect of magnetic field on VO coupling, moreover we discuss the role of excited states on the VO coupling in the ground state manifold. In the final section we summarize our results and discuss their importance to the experimental study of Si quantum dots.

\section{Theoretical Approach}\label{TA}

\subsection{Model Hamiltonian}\label{model}

Typical heterostructures for Si-based quantum computing employ either on silicon oxide or Si$_{1-x}$Ge$_x$ alloy as barrier materials.  The former tends to have relatively flat interface, while the latter often has a small miscut angle to release the strain built up in the growth process, which in turn results in discrete interface steps.  In this work we focus on the Si/SiGe heterostructures, and explore in detail the effects of the interface steps on the electronic states in a single gate-defined quantum dot.

Our model consists of a single electron in a quantum dot with an in-plane ($x$ and $y$ directions) circular confinement of a nominal radius of $\ell_0 = 10$ nm centered at the origin. The out-of-plane confinement is triangular, with a barrier height of $U_0$, and can be tuned by an applied electric field $F$.  The total Hamiltonian of the electron is $H^{(i)}=H_{xy}+H_z^{(i)}$, with superscript $^{(i)}$ indicating the number of steps at the interface within the quantum dot.

The in-plane part of the one-electron single-dot effective mass Hamiltonian takes the form,
\ben\label{Hamiltonian}
H_{xy}=-\frac{\hbar^2}{2m_t}\left( \frac{\partial^2}{\partial x^2} + \frac{\partial^2}{\partial y^2} \right)+\frac{\hbar^2}{2 m_t \ell_0^4}\left(x^2+y^2\right)
\een
where $m_t = 0.192 m_0$ is the transverse effective mass of an electron in Si.

Valley-orbit coupling originates from electron scattering at the interface, governed by $H_z$.  For a smooth interface positioned at $z_I$, with SiGe in the $z<z_I$ region while Si in the $z> z_I$ region, the out-of-plane part of the Hamiltonian is,
\ben\label{step0}
H_z^{(0)}\left(z_I\right)=-\frac{\hbar^2}{2m_z} \frac{\partial^2}{\partial (z-z_I)^2} +eF(z-z_I)+U_0 \theta (z_I-z) \,.
\een
Here $e$ is the elementary charge, $m_z = m_l = 0.98 m_0$ is the longitudinal effective mass of an electron, $F$ is the interface electric field,
and $\theta$ is the Heaviside step function.  For Si/SiGe $U_0=150$ meV.  We use an applied electric field $F=1.5$ MV/m, as is typical in the literature \cite{Saraiva_2009, Culcer_2010b}.

Within the effective mass approximation, the electron wavefunction at a conduction minimum can be written as a product of an envelope function and the underlying Bloch state,
\ben\label{wfn}
D^{(i)}_\xi({\bf{r}})=F^{(i)}({\bf{r}})u_\xi ({\bf{r}})e^{-i{\bf k_\xi} \cdot {\bf r}}
\een
Here $F^{(i)}({\bf{r}})$ is the envelope function with the superscript indicating the type of the step(s) at the interface and $u_\xi ({\bf{r}})=\sum_{\bf{K}} c_{\bf{K}}^\xi e^{i{\bf{K}}\cdot{\bf{r}}}$ are the Bloch states with $c^\xi$ the Bloch coefficients. Here $\xi= \{z,-z\}$ is the valley index, and $ {\bf k_\xi} = \pm 0.85(2\pi/a_{Si}) \hat{z}$ represents the location of the $\pm z$ band minima in the First Brillouin Zone of silicon with $a_{Si}=0.543$ nm being the lattice constant of Si. For a smooth interface the envelope function for the ground state can be further factored into into the in-plane and out-of-plane parts: $F^{(0)}(x,y,z)=\phi (x,y) \psi_I (z)$, where $\phi(x,y) =\frac{1}{\pi \sqrt{\ell_0}}e^{-\frac{x^2+y^2}{2\ell_0^2}}$ is the ground state of a two-dimensional harmonic oscillator and $\psi_I(z)$ is the modified Fang-Howard (mFH) wavefunction along the $z$ direction,
\begin{eqnarray}\label{stepfunction}
\nn \psi_I(z) & = & N z_0 e^{\frac{k_b\left(z-z_I\right)}{2}}\theta\left(z_I-z\right) \\
& + & N \left(z-z_I+z_0\right)e^{-\frac{k_{Si}\left(z-z_I\right)}{2}}\theta\left(z-z_I\right) \,,
\end{eqnarray}
where $N$ is the normalization factor.  The modified Fang-Howard wave function depends sensitively on the position of the interface $z_I$ \cite{Bastard, Culcer_2010a}, with $k_b$ determined by the interface potential, $z_0$ computed from the continuity of wavefunction about $z_I$, and $k_{Si}$ a variational parameter determined by minimizing the electron energy and dependent on the applied electric field along $z$. For our SiGe barrier, with U$_0$ = 150 meV, k$_{Si}$ = 0.98 nm$^{-1}$.
\begin{figure}[b!]
\includegraphics[width=.45\textwidth]{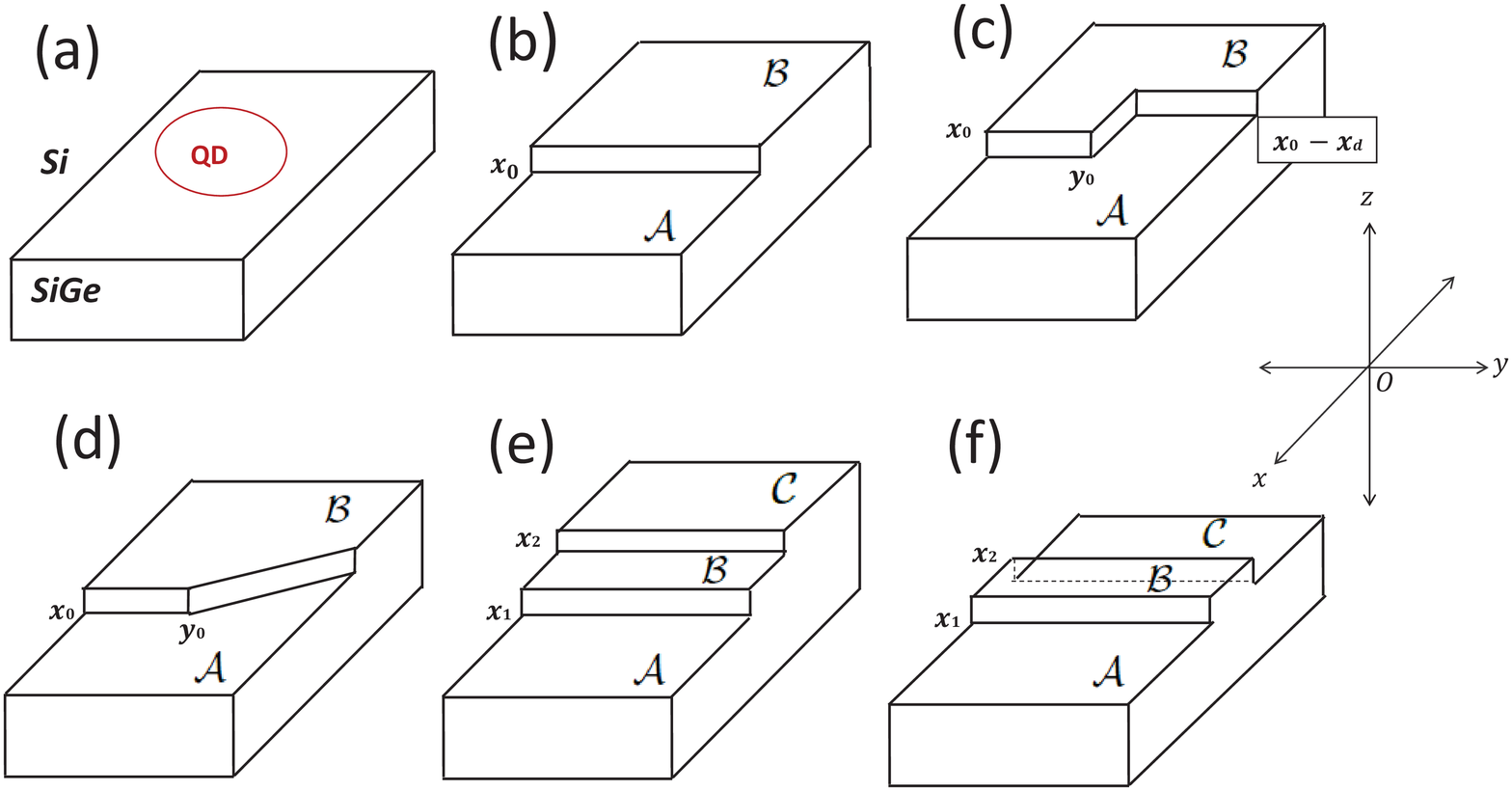}
\caption{(Color outline): Schematic representations of different SiGe alloy interfaces that are discussed in this paper.}
\label{GR}
\end{figure}

\subsection{Variational Wavefunction}

During the growth process of Si/SiGe heterostructure, interface steps are inevitable \cite{Zanenberg_RMPReview, Friesen_2007}. The step heights are an integral multiple of atomic layer between the Si atoms. For a step of one atomic layer, its height is $d=a_{Si}/4=0.136$ nm.
In the presence of an idealized straight interface step at $x = x_0$, the $z$-component of the potential is now $x$-dependent,
\ben\label{H1step}
H^{(1)}_z \left(x_0\right)=H_z^{(0)}\left(z_\mathcal{A}\right)\text{ }\theta (x_0-x)+H_z^{(0)} \left(z_\mathcal{B}\right) \text{ } \theta (x-x_0) \,,
\een
where $H_z^{(0)}\left(z_\mathcal{A}\right)$ and $H_z^{(0)}\left(z_\mathcal{B}\right)$ are the Hamiltonians given in Eq.~$\left(\ref{step0}\right)$, with smooth interfaces at $z_\mathcal{A}$ and $z_\mathcal{B}$, respectively, as shown in Fig \ref{GR}(b).

Effects of an interface step has been considered within the effective mass approximation before, for example using a smooth tilted interface approximation \cite{Friesen_2007,Friesen_2006,Friesen_2010}. Here we choose to use the Heaviside theta function to divide the two sides of a step so as to preserve the information on step location, which turns out to be a crucial factor in determining the valley-orbit coupling.

The interface step couples the in-plane and out-of-plane degrees of freedom, making an analytical solution to the electron wave function impossible.  We therefore take a variational approach in writing down the envelope function for the QD-confined electron.  The main requirement here is to match the modified Fang-Howard wave function with the correct interface position along the $z$ direction because the energy expectation value increases dramatically if there is a mismatch.  In the meantime, as an electron crosses an interface step, its $z$-direction wave function should smoothly shifts from one mFH function to another, so that the wave function remains differentiable at all points in space.  This smooth transition is achieved by stitching together the mFH functions from regions $\cal{A}$ and $\cal{B}$ (Fig.~\ref{GR}(b)) using a complimentary error function (Erfc)\cite{Erfc}. 

Consider an example of a single interface step, where we define $\psi_\mathcal{A} (z)$ as the ground eigenfunction of Hamiltonian $H_z^{(0)} \left(z_\mathcal{A}\right)$ and $\psi_\mathcal{B}(z)$ for $H_z^{(0)}\left(z_\mathcal{B}\right)$.  The total envelope function is then
\bea\label{wfn1step}
\nn F^{(1)}\left(x_0\right)=\frac{N_1}{2} \phi (x,y) \Biggl[ \psi_\mathcal{A} (z)\text{Erfc}\left( \frac{x-x_0}{L_x} \right) \\
+\psi_\mathcal{B} (z)\text{Erfc}\left( \frac{x_0-x}{L_x} \right) \Biggr] \,,
\eea
where $N_1$ is the normalization constant, and $L_x$ is the width of the error function around the step position $x_0$.  $L_x$ is a variational parameter obtained through minimizing the expectation value of the ground state energy $E^{(1)}=\left\langle F^{(1)}(r)\left| H^{(1)}(r) \right| F^{(1)}(r) \right\rangle $.  For a one-monolayer step we find $L_x = 1.5$ nm, while for a two-monolayer step $L_x = 1.0$ nm.  In Appendix \ref{Ap1} we give a more detailed discussion on $L_x$.

Interface steps are never truly in a straight line. There are always zigzags along the step as atoms diffuse on the surface during the formation of the interface \cite{step_bunching_1}.  We thus also consider two examples as shown in Fig.~\ref{GR}(c) and (d), with one having a sharp turn along $y$ while the other has a kink, with the steps on the two sides of the kink extending at different angles.  We have also considered an interface with two steps within the range of the quantum dot, as shown in Fig.~\ref{GR}(e) and (f).  We construct the wave functions in the same manner as in the case of a single-step, stitching together wave functions with different interface locations using complementary error functions.  Details of the wave functions are discussed in the Appendix \ref{Ap1}.

\section{Valley-Orbit Coupling}\label{VO_section}

The key quantity of interest to us in this study is the valley orbit coupling, which is defined as
\ben\label{VOg}
\Delta^{(i)} = \left\langle D_z^{(i)}\left| V_z^{(i)}(x,z) \right| D_{-z}^{(i)} \right\rangle \,,
\een
where $V_z^{(i)}(x,z)$ is the interface potential along the $z$ direction. For a smooth interface at $z=z_I$, $V_z^{(i)}(x,z)$ is independent of $x$ and takes the form,
\ben\label{VO-0}
V^{(0)}_z(z_I)=eF(z-z_I)+U_0\text{ } \theta (z_I-z) \,.
\een
For an interface with one straight step, the potential can be written as
\ben\label{1stepV}
 V^{(1)}_z(x_0)=V^{(0)}_z(z_\mathcal{A})\text{ } \theta (x_0-x) +V^{(0)}_z(z_\mathcal{B})\text{ }\theta (x-x_0) \,.
\een
The more irregular steps can be defined similarly, as discussed in the Appendix \ref{Ap1}.  For two steps,
\bea
\nn V^{(2)}_z(x,z)=V^{(0)}_z(z_\mathcal{A}) \text{ }\theta (x_1-x)&+&V^{(0)}_z(z_\mathcal{B}) \text{ }\theta (x-x_1)\text{ } \\
\nn \theta (x_2-x) &+&V^{(0)}_z(z_\mathcal{C})\text{ } \theta (x-x_2) \,.
\eea
Below we present our results on VO coupling for all cases using the above potentials.

\subsection{VO Coupling for a Smooth Interface}

As a benchmark for VO coupling with interface steps, and qualitative understanding on how the phase of VO coupling arises, we first examine the effect of a smooth interface. For an interface at $z=z_I$, we obtain an analytical expression of VO coupling in the absence of Umklapp processes by substituting Eq.~(\ref{wfn}) and Eq.~(\ref{VO-0}) into Eq.~(\ref{VOg}),
\ben
\Delta^{(0)}=U_0 c_0^z c_0^{-z*}  \frac{ N^2 z_0^2}{ k_b-2ik_0} e^{-2ik_0z_I} \,.
\een
Here $c_0^z$ and $c_0^{-z*}$ are the Bloch coefficients.  The magnitude and phase of the VO coupling depends on the strength of the applied electric field along the growth direction via the Fang-Howard parameter $k_{Si}$, and the location of the interface.  The latter produces a phase shift, which is a trivial global phase for a smooth interface and can be removed by assuming $z_I = 0$. However, when steps are present, the variations in this phase shift play a crucial role in determining the overall magnitude and phase of the VO coupling.

The phase shift $2k_0z_I$ arises from electron scattering between the two conduction band minima at $\pm k_0$.  With $z_I=0$ as a reference, the interface at $z_I=\pm d= \pm a_{Si}/4$ leads to a phase shift of $\pm 2 k_0 d$.  For a monolayer step, $d$ is one quarter of the Si lattice constant, so that the phase shift is $\pm 0.85 \pi$ ($\pm 153^\circ$) (the range of phase is $\left\{-\pi,\pi\right\}$). Hence, for step with height $z_I=\pm 2 d$, the relative phase is $\pm 0.30 \pi$ ($\pm 54^\circ$). This phase shift across an interface step is at the heart of how a step modifies the overall VO coupling at the interface.

\begin{figure}[t!]
\includegraphics[width=.45\textwidth]{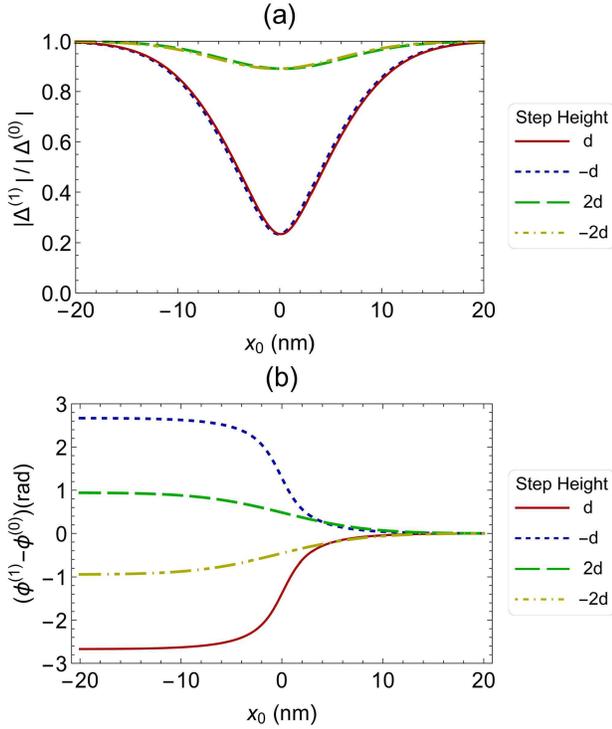}
\caption{(Color online) In (a) we have shown the ratio of magnitude of valley orbit coupling in the presence of a step to smooth interface and in (b) phase change in valley orbit coupling between interface have a step to smooth interface, as a function of position of step. Here $d=a_{Si}/4=0.543$ nm is the monolayer height of the step.}
\label{1stepa}
\end{figure}

\subsection{VO Coupling: An interface with One Step}\label{1step_sec}

Now we calculate the effect of a single interface step on the VO coupling, with particular focus on how the magnitude and phase of the VO coupling depend on the location and height of the step.

Figure \ref{1stepa}(a) shows the numerical results of the magnitude of VO coupling as a function of the step position.  Clearly, when a step is introduced, it always leads to suppression in the magnitude of the VO coupling because the complex contributions from the two sides of the step (regions $\mathcal{A}$ and $\mathcal{B}$ in Fig.~\ref{GR}(b)) have different phases \cite{Friesen_2010}.  For a monolayer step this phase difference is $0.85 \pi$.  If the step is positioned at the center of the QD, the two regions' contributions to the VO coupling are equal in magnitude but almost opposite in phase, so that they largely cancel each other out.  The resulting VO coupling has a magnitude that is only $23\%$ of the smooth-interface value, as shown in Fig.~\ref{1stepa}(a).  When the step moves away from the center of the QD, one region starts to make larger contribution to VO coupling than the other, until the step moves outside the QD, at which point the VO coupling magnitude recovers its smooth-interface value.  These results are consistent with the recent tight-binding calculations reported in Ref.~\onlinecite{Friesen_2018}.

Figure \ref{1stepa}(b) shows the phase of the VO coupling as a function of the step position. On the right side of the curve, the phase is zero (by defining that the particular interface location $z_I = 0$) and on the left side the phase is different for step up and step down case because of the different interface location, $z_I = \pm d$. The change around the center of QD is faster than the edges, which is the consequence of the varying complex contributions from each regions.
A change in the QD size along $x$ direction does not alter the general behavior of the curve in either magnitude or phase, but modifies the width of the changes around the center of the QD. A change in the $y$ dimension has no effect on VO coupling here, while a change in the $z$ dimensions brings a change in the VO coupling for a smooth interface, but does interfere with the effects of the interface step.

Under certain experimental conditions, interface steps may become bunched during the growth process\cite{step_bunching} and the resulting step height is of two atomic layers.  Now the phase change across the step is $4 k_0 d = 1.7\pi = 306^\circ$, which is equivalent to $-0.3\pi = -54^\circ$.  This phase difference is much less than the single monolayer case, so that the suppression in the magnitude of VO coupling is now less severe. The minimum value for VO coupling still occurs at the center of the QD, though the reduction is only about 10 $\%$ of the original value. The phase varies from $0$ to $\pm 54^\circ$ as shown in Fig.~\ref{1stepa}(b), with the step-up and step down cases opposite in their trends.

Considering that each contribution to the VO coupling from a particular step region is a complex number and can be represented by a two-dimensional vector, our results can be explained using a simple vector model.  Mathematically, the VO coupling defined in Eq.~(\ref{VOg}) can be expressed as follows in the presence of an interface step and neglecting the overlapping contributions from the two sides of the step (assuming complementary error function as a step function since $L_x<<\ell_0$)
\begin{widetext}
\ben
\begin{split}
 \Delta^{(1)} & \approx  \Sigma_0 \int dxdydz\text{ } e^{-2ik_0z} \left| \phi(x,y) \right|^2 \psi_\mathcal{A}(z)^2 \theta \left(x-x_0\right)  + \Sigma_0 \int  dxdydz\text{ } e^{-2ik_0z}  \left| \phi(x,y) \right|^2 \psi_\mathcal{B}(z)^2 \theta \left(x_0-x\right)  \,, \\
 &   \approx  \Delta_0 \left( \int_{-\infty}^{x_0} dx\text{ } e^{-\frac{x^2}{a^2}} \theta \left(x-x_0\right) +e^{-2ik_0\left(z_B-z_A\right)} \int_{-\infty}^{\infty} dx\text{ } e^{-\frac{x^2}{a^2}} \theta \left(x-x_0\right) \right)\,, \\
&=\Delta_\mathcal{A}+\Delta_\mathcal{B}\,.
\end{split}
\een
\end{widetext}
where $\Sigma_0=U_0 c_0^z c_0^{-z*}$. In the last expression above each term can be assigned a two-dimensional vector. The first vector represents contribution from region $\mathcal{A}$, the second vector region $\mathcal{B}$.  The magnitude of a vector is given by the electron probability in the corresponding region, while the relative direction of the vectors given by the phase $e^{-2ik_0\left(z_B-z_A\right)}$.

For the two interface regions $\cal{A}$ and $\cal{B}$ in Fig.~\ref{GR}(b)-(d), we associate each as a vectors with the following properties:
\begin{itemize}
\item The direction of each vector is fixed by the corresponding $z_I$, as $-2ik_0 z_{I}$ measured from the $x$-axis.
\item The magnitude of each vectors depends on the fraction of electron probability within the specific step region in the QD.
\item The algebraic sum of the magnitudes of all the vectors is a constant (for one step here $\left|\Delta_\mathcal{A}\right| + \left| \Delta_\mathcal{B} \right| = \left| \Delta^{(0)} \right|$), reflecting normalization of the electron wave function.
\end{itemize}
Consider the example of a step at the center of the quantum dot. Here the vectors from the two sides of the step have the same magnitude ($\left|\Delta_\mathcal{A}\right|=\left|\Delta_\mathcal{B}\right|=\left|\Delta^{(0)}\right|/2$) but are directed at different angles. The relative angle between the vectors for a monolayer step is $153^\circ$. Hence the ratio of the magnitude should be $\left|\frac{1}{2}\left(1+e^{-i0.85\pi}\right)\right|=0.23$, consistent with our numerical calculation. When the step moves away from the center , the magnitudes of the two vectors increases and decreases with the same pace and can be represented as $\left|\Delta_\mathcal{A}\right|+\delta$ and $\left|\Delta_\mathcal{B}\right|-\delta$.  Assuming the phases of the two contributions to be $0$ and $\theta$, the sum $\Delta^{(1)}$ of these two vectors/complex numbers can be expressed as
$\Delta^{(1)} = \left( \left|\Delta_\mathcal{A}\right|+\delta \right) + \left(\left|\Delta_\mathcal{B}\right|-\delta \right) e^{i\theta}$.  Its magnitude is
\ben
\left|\Delta^{(1)}\right| = \sqrt{\frac{\left|\Delta^{(0)}\right|^2}{2}+ 2\delta^2 +2\left(\frac{\left|\Delta^{(0)}\right|^2}{4} - \delta^2\right) \cos \theta}\,,
\een
which is an even function of $\delta$ (which in turn is proportional to the step location measured from the center of the dot).  $\left|\Delta^{(1)}\right|$ also only depends on $\cos \theta$, so that the sign of $\theta$ does not matter, which implies that the VO coupling for the step-up and step-down situations should have the same magnitude.  The phase of the sum is given by
\ben
\tan \phi = \frac{(\left|\Delta^{(0)}\right|-2\delta) \sin\theta}{\left|\Delta^{(0)}\right|+2\delta+\left(\left|\Delta^{(0)}\right|-2\delta\right) \cos \theta}\,,
\een
 which approaches $0$ when $\delta \rightarrow -\left|\Delta^{(0)}\right|/2$ (smooth interface with $z_I = 0$) and $\theta$ when $\delta \rightarrow\left|\Delta^{(0)}\right|/2$ (smooth interface with $z_I$ a constant of $d$ or $2d$).  Therefore, our results presented above can be almost exactly represented by the behavior of the sum vector $\Delta^{(1)}$.

\begin{figure}[h!]
\centering
\includegraphics[width=.45\textwidth]{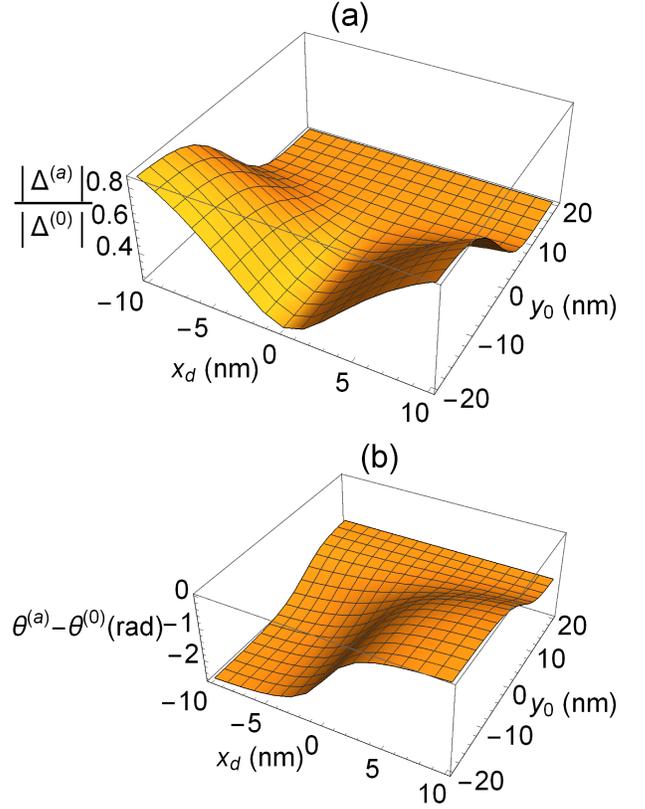}
\centering

\caption{(Color online) Three dimensional plot of the ratios of magnitudes in (a) and phase differences in (b) of VO coupling as a function of positions of the step at $y_0$ and a sharp turn of width $x_d$ in $x$ direction.}
\label{ystep}
\end{figure}

\subsection{VO Coupling: Irregular One Step}

With wave function overlap negligible across a step, the VO coupling is mostly determined by the contributions from different regions proportional to their respective areas within the QD range, as illustrated by the excellent agreement between our simple vector model and the numerical results for a single straight step above.  Within this model, the shape of the step also does not matter.  To demonstrate this point, we introduce two irregular steps, one with a straight step along $y$ direction but has a sharp zigzag turn along $x$ direction [Fig.~\ref{GR}(c)], the other with two segments that are at an angle between each other [Fig.~\ref{GR}(d)].

Figure \ref{ystep} presents the results of the zigzag turn case, with the VO coupling as functions of variables $x_d$ (the length of the $x$-direction segment) and $y_0$ (the location of the zigzag).  $y_0=-20$ nm represents the point where the zigzag is below the QD so that there is only a straight step through the QD.  This is thus similar to the straight step case.  in particular, when $x_d=0$ we get a suppressed VO coupling of $23\%$ magnitude, the same as before, and the phase is also the same as expected.  Hence, we see the similar results to one step for both magnitude and phase. At $y_0=0$ for $x_d=10$ nm, the area occupied by region $\cal{A}$ is three-quarter whereas, region $\cal{B}$ one quarter. They both will add-up with the phase factor and gives the result of approx. $50\%$ change in magnitude. This case is symmetric when region $\cal{B}$ occupies three-quarter and region $\cal{A}$ one quarter. On the other hand, when $y_0=20$ nm the change of $x_d$ does not affect the VO coupling. The changes happening outside the dimensions of the QD and we see no change in the result.

\begin{figure}[t!]
\includegraphics[width=.45\textwidth]{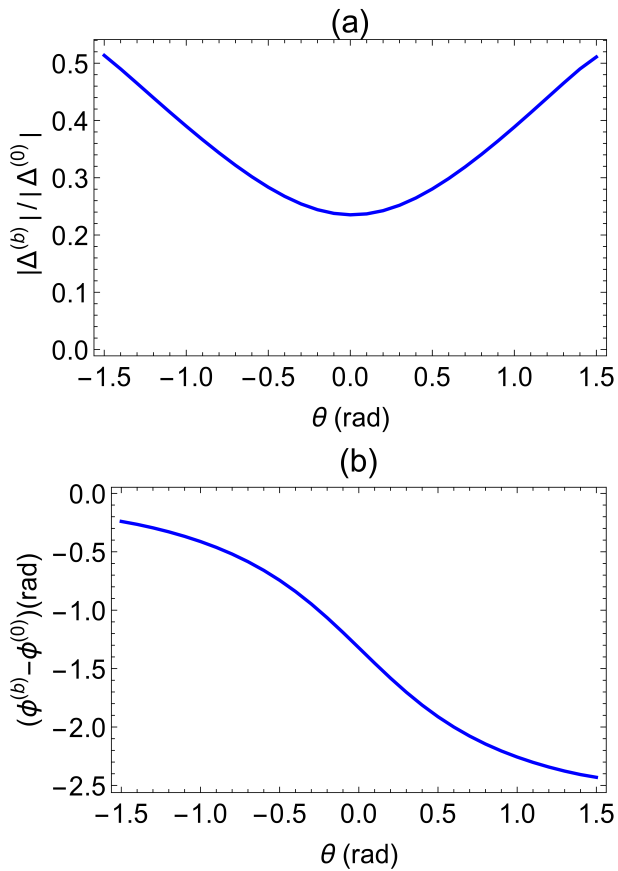}
\caption{(Color online) Magnitude (a) and phase (b) of VO coupling as a function of tilt angle ($\theta$) for geometry given in Fig. \ref{GR}(c). We fixed $x_0=0$ and $y_0=0$, i.e., at the center of the QD.}
\label{y_angle}
\centering
\end{figure}

For the angled-turn case shown in Fig. \ref{GR}(d), we take $y_0$ to be fixed at the center of the QD and study the VO coupling's dependence on the turning angle $\theta$.  For example, when $\theta = -\pi/2$, region $\cal{A}$ occupies three-fourth of the total area of the QD, while region $\cal{B}$ occupies one-fourth. The results presented in Fig.~\ref{y_angle}, for both magnitude and phase of the VO coupling, are the same as in Fig.~\ref{ystep}.  In short, the results illustrates again our point that it is the areas of each region, not the shape the boundary, that are the determining factor for the overall VO coupling.

\subsection{VO Coupling for an Interface with two Steps}\label{2steps}

When two straight steps are present on an interface, the VO coupling in a QD is the sum of contributions from three step regions labeled as $\mathcal{A}$, $\mathcal{B}$ and $\mathcal{C}$, as illustrated in Fig.~\ref{GR}(e) and (f).  These regions differ by the $z$ position of the barrier potential between Si and SiGe.  There are four possible geometrical combinations: stairs (upward or downward) and rectangular terraces (normal or inverted), as discussed in Sec.~\ref{2stepwf}. The VO couplings of stairs-up and -down cases are differentiated only through their phases, similar to the one-step case.  Without loss of generality, we focus on the stair-up and rectangular terrace cases in the following calculations and discussions.

\subsubsection{VO coupling for Stairs}\label{Stairs}

\begin{figure}[b!]
\includegraphics[width=.45\textwidth]{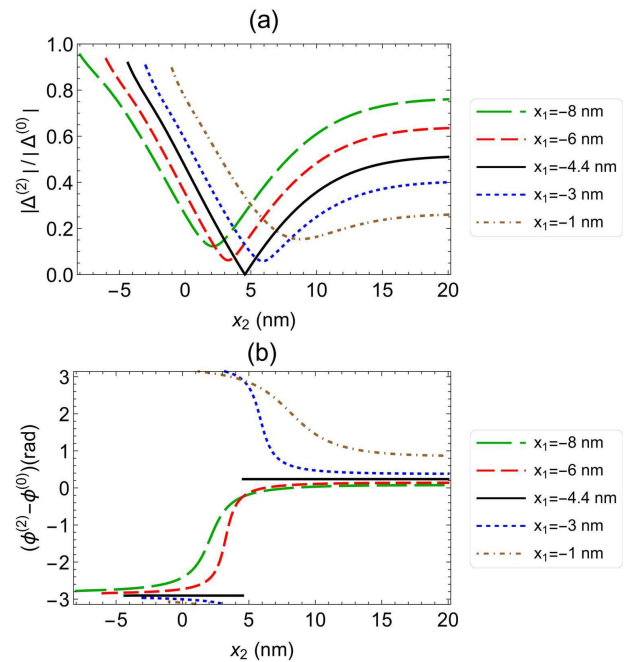}
\caption{(Color online) Magnitude ratios and phase differences for stair case with each step of height one monolayer as a function of second step. We use different fixed position of the the first step and the position of step two begins from first step location.}
\label{StUp_1}
\end{figure}

Figure \ref{StUp_1} shows the results of both magnitude and phase of the VO couplings as functions of the locations of the two steps in a stair configuration.  The most prominent feature for the magnitude of VO coupling here is that it can actually reach zero for a specific set of step locations, when the steps are at $x_1 = x_{10} = -4.4$ nm and $x_2 = x_{20} = -x_{10} = 4.4$ nm (recall that the radius of the QD is 10 nm).  While the chance of getting such a step configuration is small in a real QD, the VO coupling could be strongly suppressed if the two steps are close to this configuration.

Physically, at this particular step configuration, the contributions from each region towards the VO coupling cancel one another because of their phase differences. Within our vector model, if we let $\Delta_\mathcal{A}=\left|\Delta_\mathcal{A}\right| e^{-i\theta}$, $\Delta_\mathcal{B} = \left|\Delta_\mathcal{B}\right|$, and $\Delta_\mathcal{C} = \left|\Delta_\mathcal{C}\right| e^{i\theta}$ represent the contributions from each of the step regions, with $\theta = 0.85\pi$, the condition for a complete cancellation of the VO coupling in the QD is

\begin{eqnarray}
\left|\Delta_\mathcal{A}\right| & = & \left|\Delta_\mathcal{C}\right| = \frac{\left|\Delta^{(0)}\right|}{2(1-\cos\theta)} \\
\left|\Delta_\mathcal{B}\right| & = & -\frac{\left|\Delta^{(0)}\right|\cos\theta}{1-\cos\theta} \,.
\end{eqnarray}
With the magnitude of each contribution linearly proportional to the electron probability in the respective step region, which is roughly proportional the area of the region within the QD.  Quantitatively, the middle step should occupy about 47\% of the electron probability in the QD, while the other two steps split the rest.

\begin{figure}[t!]
\includegraphics[width=.45\textwidth]{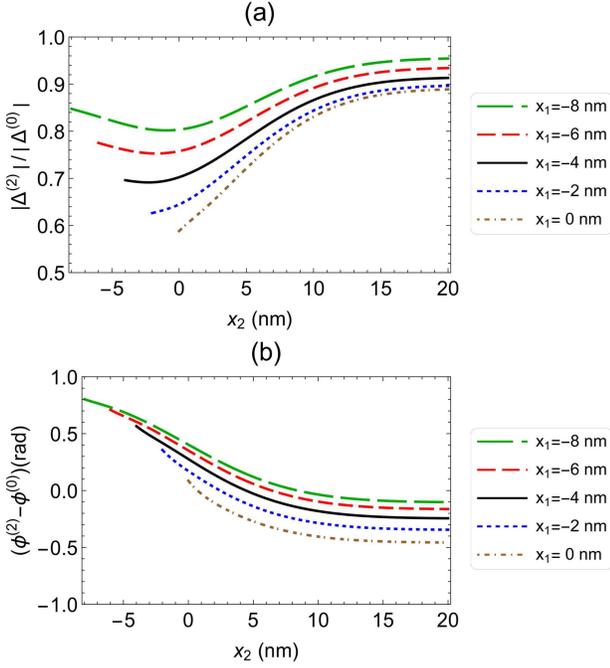}

\caption{(Color online) Results of magnitude ratios and phase differences for stair case with each step of height $2d$ for several fixed locations of first step as a function of second step position.}
\label{StUp_2}
\end{figure}

The phase of the VO coupling is ill defined when its magnitude vanishes, and could have a discontinuity as we change the step configurations.  This is indeed the case as shown by the black solid line in Fig.~\ref{StUp_1}(b), which represents the configurations with $x_1 = x_{10}$ while $x_2$ is varied around $x_{20}$.  Somewhat surprisingly, in this case the phase has only two values that differ by $\pi$.  This feature can again be explained within our vector model.  Qualitatively, when $x_1$ is fixed at $x_{10}$, the total vector changes only its magnitude but not its angle when $x_2$ is varied.  When $x_2$ sweeps past $x_{20}$, the vector simply flips its direction, so that the corresponding complex number changes its phase by $\pi$.  For a more detailed discussion please see Appendix \ref{Ap2}.

Doubling the height of each step changes the phases of regions $\mathcal{A}$ and $\mathcal{C}$ to $\pm 4k_0d = \mp 0.3 \pi$ (assuming the phase of region $\mathcal{B}$ contribution is $0$).  Now the overall VO coupling cannot vanish anymore, and generally has much larger magnitude than the case of two single-monolayer steps.  The minimum magnitude occurs when the two steps merges at the center of the QD with a height of $4d$, with the value of $\left|\frac{1}{2}\left(1+e^{-4i0.85\pi}\right)\right|=0.59$. As the second step moves outside the dot the magnitude of VO coupling gets the value of one step with height $2d$. The phase of the VO coupling smoothly shifts from the case of a step of height $4d$ to a height of $2d$.  In term of the vector notation, here the magnitude of total VO coupling never goes to zero, and there is no discontinuity in its phase.

\subsubsection{VO Coupling in Rectangular Terraces}

\begin{figure}[b!]
\includegraphics[width=.45\textwidth]{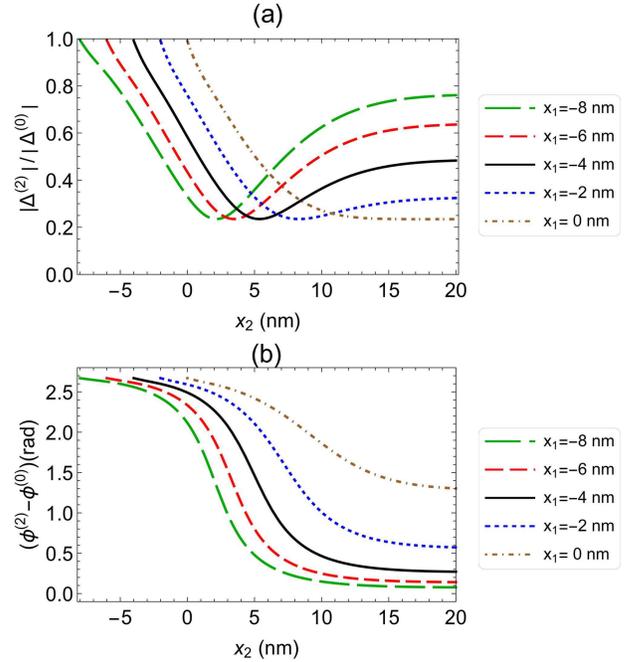}
\caption{(Color online) Magnitude ratios and phase differences for rectangular terrace of height one monolayer, $d$, as a function of second step. We use different fixed position of the the first step and the position of step two begins from first step location. }
\label{Te_1}
\end{figure}

When a rectangular terrace is present on the interface (a step up followed by a step down), the magnitude of VO coupling has a non-zero minimum value just like in the single-step case.  Furthermore, this minimum value is identical to the value obtained in the single-step case as well.  In Fig. \ref{Te_1}(a) we show some numerical results for different terrace configurations, with five different locations of first step on the left side of the quantum dot. The second step position always satisfies $x_2 > x_1$.  Each of the curve starts from 1 when the two steps are on top of each other so that the terrace is absent. As the second step moves away from the first step, the value of the magnitude of VO coupling drops to a minimum value and then goes back up again.  No matter where $x_1$ is (as long as $x_1 < 0$), the minimum value of the VO coupling remains the same.  As for the phase of the VO coupling, for each $x_1$ the phase starts from zero when both steps are at the same point and there is no terrace.  The contribution from the terrace increases as $x_2$ moves away from $x_1$ and the terrace area becomes larger. Finally, when the second step moves outside the QD we recover the single-step case (with step location at $x_1$) as shown in Fig.~\ref{1stepa}(b).

\begin{figure}[t!]
\includegraphics[width=.45\textwidth]{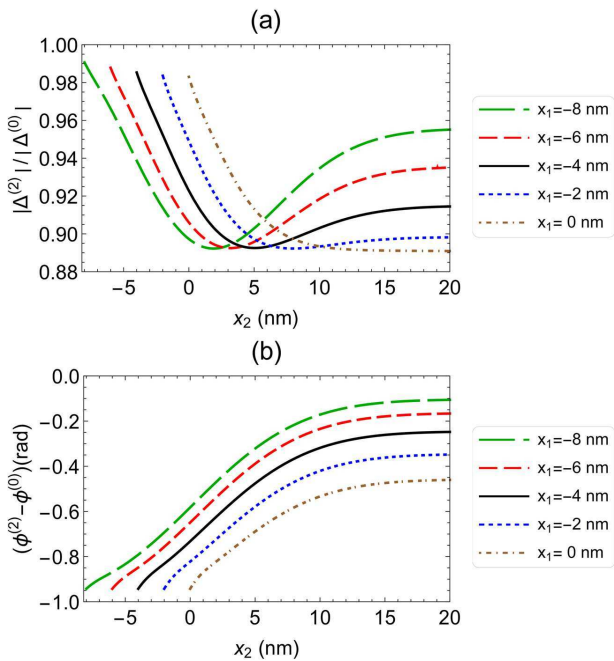}
\caption{(Color online)  Results of magnitude ratios and phase differences for rectangular terrace of height $2d$ for several fixed locations of first step as a function of second step position. }
\label{Te_2}
\end{figure}

These results can be interpreted by our vector model straightforwardly. As in the case of a stair, here the contributions to the VO coupling again comes from three regions, labeled as $\mathcal{A}$, $\mathcal{B}$, and $\mathcal{C}$.  What is different here is that regions $\mathcal{A}$ and $\mathcal{C}$ have the same location for the interface, so that the vectors representing these two regions are aligned and can be combined into a single vector, whereas the contribution from region $\mathcal{B}$ is at an angle $0.85\pi$ from those of $\mathcal{A}$ and $\mathcal{C}$.  Clearly, the sum of these three vectors (practically two if $\mathcal{A}$ and $\mathcal{C}$ are combined) would never vanish.  Writing the total VO coupling as $\Delta^{(2)} = \Delta_\mathcal{A} + \Delta_\mathcal{B} + \Delta_\mathcal{C} = \left|\Delta_\mathcal{B}\right| + \left(\left|\Delta_\mathcal{A}\right| + \left|\Delta_\mathcal{C}\right|\right) e^{-i\theta}$, and knowing that $\left|\Delta_\mathcal{A}\right| + \left|\Delta_\mathcal{B}\right| + \left|\Delta_\mathcal{C}\right| = \left|\Delta^{(0)}\right|$ is a constant determined by the QD area, we can easily find that the minimum of $\left|\Delta^{(2)}\right|$ happens when $\left|\Delta_\mathcal{B}\right| = \frac{1}{2} \left|\Delta^{(0)}\right|$ irrespective of what the split is between regions $\mathcal{A}$ and $\mathcal{C}$, and the value of the minimum is $\left|\Delta^{(2)}\right|\sim$ 0.23 $\left|\Delta_0\right|$, the same as in the single-step case.  Again, from the perspective of this model, the key factor is the electron probability distribution among the different step regions within the QD, not where exactly the terrace is.

If the terrace step height is $2d$ (two atomic monolayer), the magnitude of the VO coupling follows the same trend as above, albeit with a smaller overall magnitude in modification, consistent with what we observed in the single-step case.  The phase also follows the same trend as in the single-step case (the $-2$d dot-dashed curve in Fig.~2), as would be predicted by our vector model.

Our results on two-step interfaces show that both the magnitude and phase of the valley-orbit coupling in a Si quantum dot are dependent on the configurations of the steps. In particular, when the two steps are in a stair formation, the VO coupling can be completely suppressed for a specific combination of step locations.  However, digging beneath the surface, our effective mass model also indicates that it is the electron probabilities in each step region within the QD that determines the behavior of the overall VO coupling. For example for the cases of terraces, the VO coupling would be minimized as long as half of the electron probability is within the terrace, no matter exactly where the terrace is within the QD's geometric shape.

\section{Effects of Magnetic Field and Higher Orbital States on the Ground State Valley Orbit Coupling}\label{B_section}

So far in this study our key observation is that in the case of a relatively smooth interface with one or two steps, the crucial factor in determining the magnitude and phase of the VO coupling is the area/electron probability of each step (with a particular interface location $z_I$) within a quantum dot.  There is no direct contribution from the step edge within our effective mass model.

\begin{figure}[b!]
\includegraphics[width=.45\textwidth]{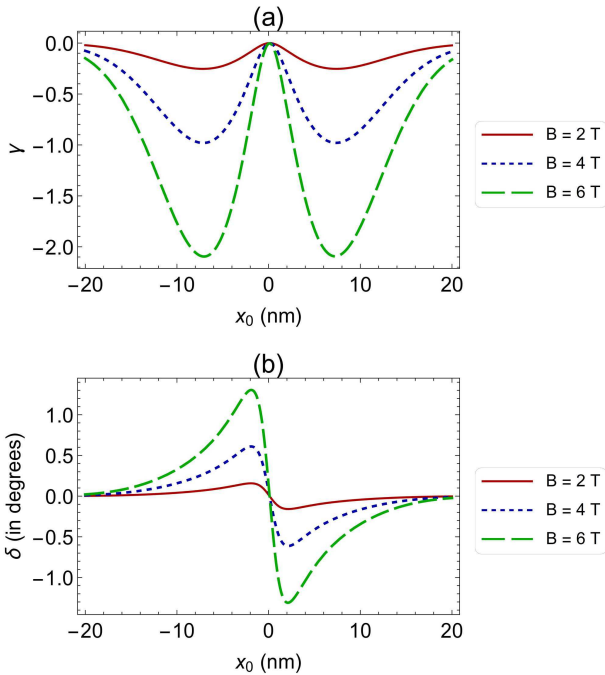}
\caption{(Color online) In (a) has shown the percentage difference in $\left|VO\right|$ and phase difference in VO given in Eqs. \ref{B_M} and \ref{B_P}, respectively, for three different values of magnetic field as a function of step location. }
\label{B1s}
\end{figure}

When a magnetic field is applied along the growth direction of a two-dimensional quantum dot, it causes a stronger confinement and a reduction in the Fock-Darwin radius of the electron wave function (i.e. the characteristic length of the electron wave function) \cite{Hu_PRA2000},
\ben
\ell (B)=\ell_0 \left[1+\frac{1}{4}\left(\frac{ \ell_0}{\ell_B}\right)^4\right]^{-\frac{1}{4}} \,,
\een
where $\ell_0$ is the radius of the ground Fock-Darwin electron state at $B=0$, and $\ell_B=\sqrt{\frac{\hbar}{eB}}$ is the magnetic length for the applied magnetic field along the $z$ direction.  Such a change in the size of a QD could theoretically shift the position of a step relative to the center of the QD, therefore leading to changes in the VO coupling.  However, since SiQD radius is generally quite small (we have used a nominal value of 10 nm in this study), this magnetic confinement effect turns out to be negligible.
Benchmarked against the zero-field results, the quantities of interest here are the percentage difference in the magnitude of VO coupling and overall phase shift,
\ben\label{B_M}
\gamma = \frac{\left|\Delta^{(1)}\left(B\right)\right|-\left|\Delta^{(1)}\left(B=0\right)\right|}{\left|\Delta^{(0)}\left(B=0\right)\right| }\times 100 \% \,,
\een
\ben\label{B_P}
\delta = \phi^{(1)}\left(B\right)-\phi^{(1)}\left(B=0\right) \,.
\een
In Fig.~\ref{B1s} (a), we indeed observe a very small change in VO coupling as a function of the magnetic field, in the presence of a single step  because of the relative small change in the confinement radius: the largest change is 4.6\%, at $B=6$ T. There is no change in VO coupling if the step is at the center of the QD because the magnetic field does not shift the electron wavefunction.  The phase of the VO coupling has a similarly very weak dependence on the magnetic field, as shown in Fig.~\ref{B1s}(b).  As we mentioned before, these weak dependence on the applied magnetic field is to be expected.  Indeed, given a 10 nm zero-field radius for the electron ground state, we have total confinement radius for different values of B field as $\ell\left(\text{2T}\right)=9.94$ nm, $\ell\left(\text{4T}\right)=9.78$ nm and $\ell\left(\text{6T}\right)=9.54$ nm.  Even at 6 T, the confinement length is still only modified by 5\%.  It is therefore not surprising that the field does not cause the step location to change within the QD, so that VO coupling is not affected significantly by a magnetic field.

Our results so far indicate that the determining factor in the calculation of the VO coupling in the presence of interface steps is the electron probability in each step region within the QD.  With our calculations based on a variational ground-state wave function, it is important to establish its validity by clarifying whether the steps cause significant scattering between the ground and excited orbital states.

\begin{figure}[t!]
\centering
\includegraphics[width=.45\textwidth]{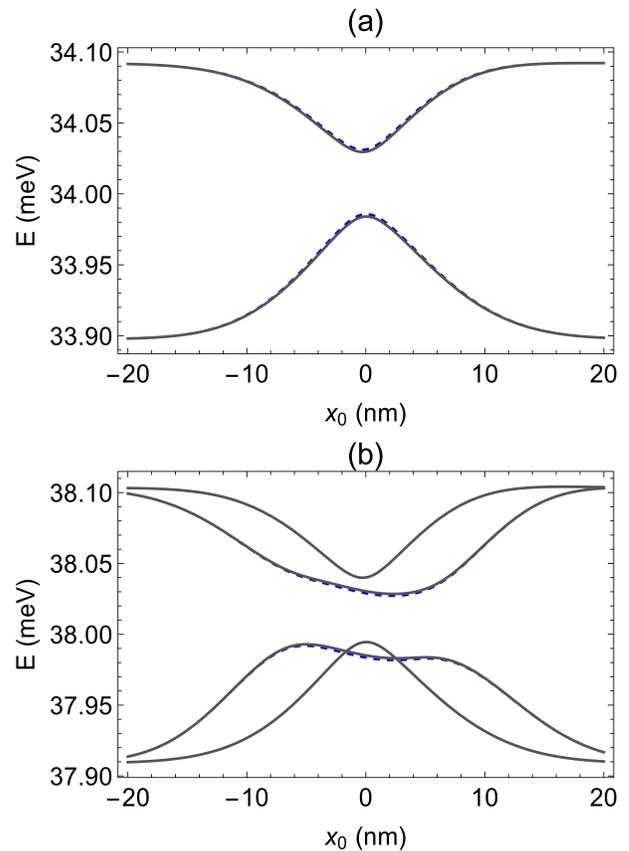}
\centering

\caption{(Color online) The ground and the first excited state energy spectrum as a function of step position are shown in (a) and (b), respectively. The solid line curves represents the results which includes the coupling between $s$ and $p$ states and dashed line approximate results with the exclusion of coupling between$s$ and $p$ states. }
\label{pstate}
\end{figure}

This can be done by introducing the higher-energy $p$ states in our calculations with a single interface step.  Now the low-energy Hilbert subspace comprises of the ground and a two-fold degenerate excited orbital states,\cite{QM} in addition to the valleys.
The basis are $\left\{D_{z,s}^{(1)},D_{-z,s}^{(1)},D_{z,p_x}^{(1)},D_{-z,p_x}^{(1)},D_{z,p_y}^{(1)},D_{-z,p_y}^{(1)}\right\}$.
Qualitatively, the presence of a step on the interface breaks the symmetry of the Hamiltonian along the $x$ direction, resulting in a coupling between the ground and the excited $p_x$ state.
\begin{figure}[h!]
\centering
\includegraphics[width=.45\textwidth]{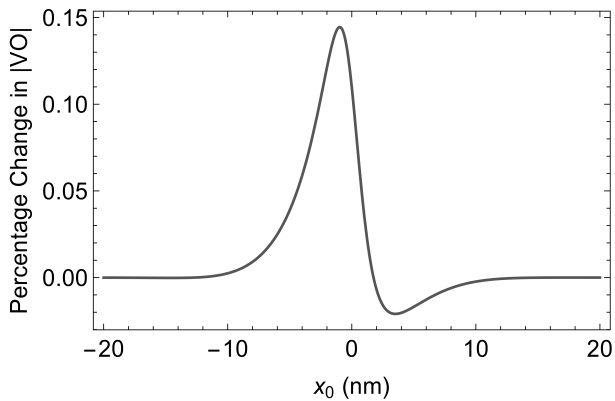}
\centering
\caption{(Color online) Percentage difference in VO coupling for ground state in the presence of $p$ as a function of step position. }
\label{pdiff}
\end{figure}

The effective Hamiltonian within the $s-p$ subspace can be represented as
\[
H
=
\begin{bmatrix}
    H_{ss} & H_{sp} \\
    H_{ps} & H_{pp}
\end{bmatrix} \,.
\]
Here the blocks $H_{ss}$ and $H_{pp}$ represent the ground and first excited orbital states manifolds (assuming valley splitting is much smaller than the orbital excitation energy) and $H_{sp}$ and $H_{ps}$ are the coupling between the ground and excited states. We diagonalize this matrix and plot the energy spectrum in Fig.~\ref{pstate}. We have also included results when assuming $H_{ps}=H_{sp}=0$, and observe a very small change as shown by the dotted lines in Fig.~\ref{pstate}.  In Fig.~\ref{pdiff} we plot the percentage change in the VO coupling in the presence of excited state as compared to ground-state-only results.  Clearly, the contribution from the excited states on the VO coupling in ground state is negligible, and our ground-state-only calculation above is justified.

\section{Conclusion}

In conclusion, we have calculated the Valley-Orbit coupling in a Si/SiGe quantum dot in the presence of interface steps.  We employ a variational approach within the effective mass approximation for this calculation.  Our results show that the presence of interface steps could lead to significant suppression of the magnitude of the VO coupling, and cause large phase shift (up to $\pi$) to the VO coupling. Our results can be explained by the assumption that the overall VO coupling is the sum of contributions from individual step surfaces.  The phase of an individual contribution is determined by the interface position of the step along the growth direction, while its magnitude is determined by the electron probability on the particular step enclosed within the QD.  This model can be visualized using vectors to represent the individual step surface contributions, with the sum of the magnitude of the vectors a constant and the direction of each vector fixed.  We have also explored the effects of an external magnetic field and the excited orbit states, and find both to have only minor effects.  Our results on the magnitude of the VO coupling are consistent with results reported in the literature, while our results on the phase of the VO coupling should be a useful guideline when exploring variations in the interdot tunneling and exchange coupling in double and multiple quantum dots.

\acknowledgments
We thank Dimitrie Culcer, Xinyu Zhao, Luke Pendo, John Truong and Zongye Wang for useful comments and discussions. This work is partially supported by US ARO through grant W911NF1710257.
\break

\appendix

\section{Variational Wave functions}\label{Ap1}

\subsection{How to determine $L_x$}

As discussed in the main text, in the presence of an interface step, we assume that the growth direction electron wave function on each step surface takes the modified Fang-Howard form.  To ensure that the wave function is continuous across a step, we stitch the two modified Fang-Howard functions together using the complementary error function:
\ben
\nn \text{Erfc}\left(\beta\right)=\frac{2}{\sqrt{\pi}}\int_{-\infty}^{\beta}e^{-t^2}\text{ }dt \,,
\een
where $\beta=\frac{x-x_0}{L_x}$, $x_0$ is the position of the step in the $x$ direction, and $L_x$ represents the width of the error function.  $L_x$ thus dictates how smooth the wave function transition is.  We treat $L_x$ as a variational parameter and calculate it by minimizing the energy of the ground state:
\bea
\nn E^{(1)}&=&\left\langle F^{(1)}(r)\left| H^{(1)}(r) \right| F^{(1)}(r) \right\rangle \,.
\eea
\begin{figure}[b!]
\includegraphics[width=.45\textwidth]{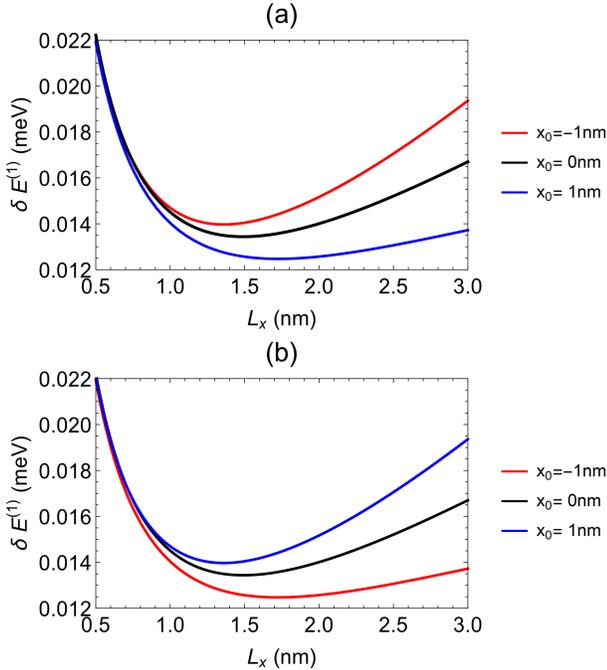}
\caption{(Color online) Change in the expectation value of energy $ \left(E^{(1)}-E^{(0)}\right)$ in the presence of (a) step up and (b) step down as a function of variational parameter, the width of complementary error function $L_x$. We consider three locations of the step close to the center of the QD.}
\label{min1step}
\end{figure}
The expectation value here depends on the location and height of the step as well as the width of the error function.  For a given height of the step, the optimal $L_x$ varies with the position of the step.  In our calculation reported in this paper, we choose three values of $x_0$ near the center of the QD, calculate the optimal value of $L_x$, and then average over them.  For the values of $x_0$ used in Fig.~\ref{min1step}, we obtain $L_x=1.5$ nm.  Notice that while the value of $L_x$ is important in calculating the total energy of a state, it has no significant impact on the calculation of the VO coupling as we discuss below.

\begin{figure}[h!]
\includegraphics[width=.45\textwidth]{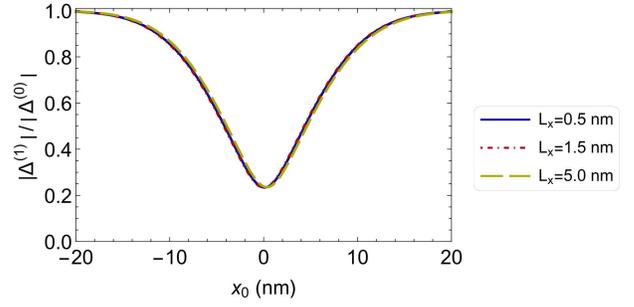}
\caption{Effect of $L_x$ on VO coupling. }
\label{App1}
\end{figure}

We observe in Fig.~\ref{min1step} that the variation in energy for the one step case is small if we vary the $L_x$ within 1 to 2 nm, whereas outside this range the change in energy is more dramatic. Most importantly, we find that for the one-step case the value of $L_x$ (e.g. 0.5 nm and 5 nm respectively) has minimal effect on the VO coupling, as shown in Fig.~\ref{App1}. In other words, the approximate nature of our choice of $L_x$ does not affect the objective of evaluating the VO coupling.

When excited states are included in our calculation, we adopt the same approach in stitching together the growth-direction wave functions across a step.  For these states the optimal value of $L_x$ is not necessarily the same as that for the ground state.
Nevertheless, we adopt the same $L_x$ for the $p$ states, and observe through numerical calculation that their effect on the ground state VO coupling is less than 0.15 percent. It is thus safe to neglect the effects of the higher-energy orbital states, as we have done in the calculations presented in the main text.

\begin{figure}[b!]
\includegraphics[width=.45\textwidth]{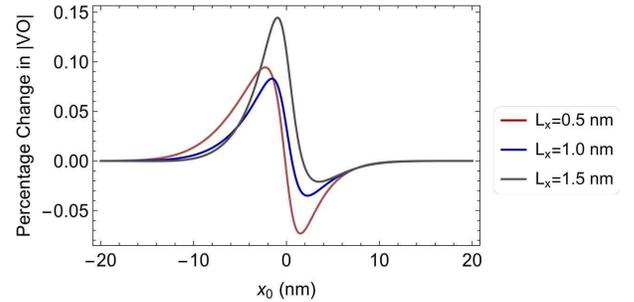}
\caption{(Color online) Percentage difference in VO coupling in the ground state in the presence of $p$ to when there is only $s$ state for different values of $L_x$. }
\label{App2}
\end{figure}

\subsection{Variational Wavefunction: Irregular One Step}\label{wyd}

On a realistic interface, a step is never a straight line.  To model irregular shapes of an interface step, we consider two examples as shown in Fig.~\ref{GR}(d) and \ref{GR}(e), with one having a sharp zigzag turn along $y$ direction and the other having two segments at an angle with respect to each other. The Hamiltonian for the two cases are,
\ben
H^{\left(\textsl{a}/\textsl{b}\right)}_z=H_z^{(1)}\left(x_0\right)\text{ }\theta (y_0-y)+H_z^{(1)}\left(x_0+x^{\left(\textsl{a}/\textsl{b}\right)}\right)\text{ }\theta (y-y_0)
\een
where $H_z^{(1)}\left(x_0\right)$ is given in Eq.~(\ref{H1step}) and the corresponding wavefunction is given as,
\bea\label{wfn2step}
\nn F^{\left(\textsl{a}/\textsl{b}\right)}\left(y_0\right)=\frac{1}{2}\frac{N_{\left(\textsl{a}/\textsl{b}\right)}}{N_1}\Bigg[ F^{(1)}\left(x_0 \right) \text{Erfc}\left( \frac{y-y_0}{L_y} \right)\\
 + F^{(1)}\left(x_0 + x^{\left(\textsl{a}/\textsl{b}\right)} \right) \text{Erfc}\left( \frac{y_0-y}{L_y} \right)  \Bigg]
\eea
where $N_{\left(\textsl{a}/\textsl{b}\right)}$ are the normalization constants, and $F^{(1)}$ is given in Eq.~(\ref{wfn1step}).  The parameters $x^{(\textsl{a})}=x_d$ and $x^{(\textsl{b})}=\left(y-y_0\right)\cot\left(\theta \right)$ define the step segment along $y$ direction and the slanted step edge in the $xy$-plane, respectively.  Compared to the case of a single straight step, here there is an additional parameter in the form of the width of the error function along $y$ direction, $L_y$.  We choose $L_y=L_x$, considering that the value of $L_x$ has a negligible effect on the VO coupling.

\subsection{Variational Wavefunction: Two Steps}\label{2stepwf}

In this Appendix we extend our formalism to the case of two parallel straight steps along $y$-direction.  Defining the locations of the two steps at $x_1$ and $x_2$, we can now divide the interface into three regions $\mathcal{A}$, $\mathcal{B}$ and $\mathcal{C}$ with respective interface positions at $z=z_\mathcal{A}$, $z=z_\mathcal{B}$ and $z=z_\mathcal{C}$, as illustrated in Fig.~\ref{GR}. Based on the locations of each regions, there are four possible configurations: the steps can form either upward or downward stairs, or rectangular terraces that can be either a bump or a dip. The Hamiltonian along the $z$ direction for both stair and terrace configurations can be written as,
\bea
\nn H^{(2)}_z=H_z^{(0)}\left(z_\mathcal{A}\right) \theta \left(x_1-x\right)+H_z^{(0)}\left(z_\mathcal{B}\right)  \theta \left(x-x_1\right)\times\\
\nn \theta \left(x_2-x\right)+H_z^{(0)}\left(z_\mathcal{C}\right)  \theta \left(x-x_2\right)\ \,,\
\eea
where $H_z^{(0)}\left(z_\mathcal{A}\right)$, $H_z^{(0)}\left(z_\mathcal{B}\right)$ and $H_z^{(0)}\left(z_\mathcal{C}\right)$ the Hamiltonians for regions $\mathcal{A}$, $\mathcal{B}$ and $\mathcal{C}$, with wavefunctions $\psi_\mathcal{A}(z)$, $\psi_\mathcal{B} (z)$ and $\psi_\mathcal{C} (z)$, respectively. Following the same procedure for the one-step case, the total wavefunction can now be written as,
\bea
\nn F^{(2)}\left(x,y,z\right)=\frac{N_2}{2} \phi (x,y)\Biggl[ \psi_\mathcal{A} (z)\text{ Erfc}\left( \frac{x-x_1}{L_x} \right) \\
\nn +\frac{1}{2}\psi_\mathcal{B} (z)\text{ Erfc}\left( \frac{x_1-x}{L_x} \right) \text{ Erfc}\left( \frac{x-x_2}{L_x} \right) \\
+\psi_\mathcal{C} (z)\text{ Erfc}\left( \frac{x_2-x}{L_x} \right) \Biggr]
\eea
Here $N_2$ is the normalization constant, which depends on the locations of the steps at $x_1$ and $x_2$, and the width $L_x$ of the complementary error function. Since the value of $L_x$ is much smaller than the dimensions of the QD and does not affect the VO coupling strongly, we assume that the widths of the complementary error functions for the two steps are the same. We use the value of $L_x$ ($=1.5$ nm), same as in the single-step case.

\section{Vector Model for VO coupling}\label{Ap2}

As we discussed in the main text, our results on valley orbit coupling can be explained in terms of vectors representing contributions from each step region. Here we provide a more detailed discussion of the vector model.  In the following we label individual vectors in terms of the step regions they represent.  After summing over all contributions, the resultant vector $\mathcal{R}$ corresponds to the total VO coupling.  For illustration we consider the cases of a single step, a two-step stair, and a rectangular terrace.

\begin{figure}[h!]
\includegraphics[width=.45\textwidth]{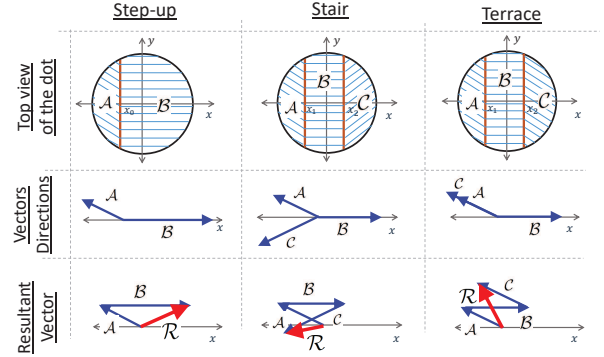}
\caption{(Color online) Vector representation of the VO coupling for a single step, a two step stair, and a rectangular terrace. We show the top view of the quantum dot in the first row, the vector representation of each shaded regions and their resultant sum are shown in the second and third rows, respectively.} \label{Ve_1}

\end{figure}

In the case of a single step, the step divides the QD into two regions labeled as $\mathcal{A}$ and $\mathcal{B}$.  A top view with step position $x_0$ is shown in Fig.~\ref{Ve_1}. The directions of the two vectors $\mathcal{A}$ and $\mathcal{B}$ are $0.85 \pi$ and $0$ from the $x$-axis, respectively, as we choose region $\mathcal{B}$ to have $z_B = 0$ so that its contribution to VO coupling is real. Their magnitudes are proportional to the electron probability within the respective areas within the quantum dot. Clearly, the magnitude and direction of the resultant vector $\mathcal{R}$  depends on the location of the step. The magnitude of $\mathcal{R}$ reaches its minimum when the step location is at the center of the dot, so that vectors $\mathcal{A}$ and $\mathcal{B}$ are of equal length. The resultant vector $\mathcal{R}$ has a finite minimum magnitude and its direction varies from direction vector $\mathcal{B}$ to vector $\mathcal{A}$ as the step moves from left to right as shown in Fig.~\ref{Ve_1}.
The presence of a second step divides the quantum dot into three regions, as illustrated in Fig.~\ref{Ve_1}. Here we consider two geometrical configurations, either two-stairs-up or a rectangular terrace. In both cases the directions of the vectors representing regions $\mathcal{A}$ and $\mathcal{B}$ are at $0.85 \pi$ and $0$ from the $x$-axis, respectively, whereas direction of $\mathcal{C}$ is $-0.85 \pi$ for stairs up case and $0.85 \pi$ for the rectangular terrace. The different direction of the third vector for the stair-up case means that the resultant vector $\mathcal{R}$ can cancel out completely. For a rectangular terrace, on the other hand, vectors $\mathcal{A}$ and $\mathcal{C}$ are parallel, so that the resultant vector $\mathcal{R}$ behaves in the same way as the case of a single step even though regions $\mathcal{A}$ and $\mathcal{C}$ are separated by $\mathcal{B}$ physically.  Like in the one-step case, the minimum value of the VO coupling (or the minimum magnitude of vector $\mathcal{R}$) is reached when the electron probability splits equally between region $\mathcal{B}$ and the collective of regions $\mathcal{A}$ and $\mathcal{C}$, and region $\mathcal{B}$ does not necessarily lie at the center of the QD.  The direction of the resultant vector $\mathcal{R}$ for both stairs-up and terrace cases varies with the change in the contributions from vectors $\mathcal{A}$, $\mathcal{B}$ and $\mathcal{C}$.  A final determination requires a more quantitative evaluation of the individual vectors, as we discussed in the main text.

\end{document}